\documentclass[conference]{IEEEtran}

\usepackage{graphicx}
\usepackage{epstopdf}
\usepackage{algpseudocode}
\usepackage{algorithm}

\begin{document}
%
\title{Cloud Enabled Emergency Navigation Using Faster-than-real-time Simulation}


\author{Huibo Bi and Erol Gelenbe\IEEEmembership{~Fellow,~IEEE}\\
Intelligent Systems and Networks Group\\
Dept. of Electrical and Electronic Engineering, Imperial College, London SW7 2AZ, UK\\
Email: \{huibo.bi12, e.gelenbe\}@imperial.ac.uk}


%


\maketitle

\begin{abstract}
State-of-the-art emergency navigation approaches are designed to evacuate civilians during a disaster based on real-time decisions using a pre-defined algorithm and live sensory data. Hence, casualties caused by the poor decisions and guidance are only apparent at the end of the evacuation process and cannot then be remedied. Previous research shows that the performance of routing algorithms for evacuation purposes are sensitive to the initial distribution of evacuees, the occupancy levels, the type of disaster and its as well its locations. Thus an algorithm that performs well in one scenario may achieve bad results in another scenario. This problem is especially serious in heuristic-based routing algorithms for evacuees where results are affected by the choice of certain parameters. Therefore, this paper proposes a simulation-based evacuee routing algorithm that optimises evacuation by making use of the high computational power of cloud servers. Rather than guiding evacuees with a predetermined routing algorithm, a robust Cognitive Packet Network based algorithm is first evaluated via a cloud-based simulator in a faster-than-real-time manner, and any "simulated casualties" are then re-routed using a variant of Dijkstra’s algorithm to obtain new safe paths for them to exits. This approach can be iterated as long as corrective action is still possible.
\end{abstract}



%
\IEEEpeerreviewmaketitle

\section{Introduction}

The increasing concentration of human populations in urbanised societies has aggravated the difficulty of evacuation planning in disasters owing to the insufficient resources for sudden massive demands and destructive crowd behaviours of evacuees. With the fast development of information and communications technology (ICT), emergency evacuation has experienced a few stages: human experience driven, in-situ wireless sensor network based navigation systems driven and cloud based navigation system driven. Accompanying this tendency, various emergency navigation algorithms have been presented to resolve solutions for this highly dynamic and complex transshipment problem. However, a single algorithm cannot ensure the optimal solution for scenarios with different conditions. For instance, Dijkstra's shortest path algorithm can only achieve good performance in scenarios with low occupancy rates because it tends to guide all the evacuees to the optimal path and causes severe congestion in densely-populated environments. Moreover, the setting of certain parameters in heuristic algorithms has a considerable impact on the simulation results and the appropriate setting for one scenario may not suit other scenarios. Therefore, it is difficult for a algorithm to achieve optimal solutions in different scenarios with varying initial conditions such as occupancy rate, distribution of evacuees and disaster source location. To solve this problem, in this paper, we propose a simulation based algorithm to optimise an evacuation process by performing faster-then-real-time evaluation in cloud servers and customise new routes for fatalities in the simulation.

The remainder of this paper is organised as follows. Section \ref{related} reviews the literature relevant to our work. Then we describe the framework of the cloud enabled emergency navigation system in Section \ref{system} and the related algorithm in Section \ref{algorithms}. The simulation models and assumptions are introduced in Section \ref{simulation} and the experimental results are presented in Section \ref{results}. Finally, we draw conclusions in Section \ref{conclusion}.

\section{Related Work}
\label{related}

Emergency navigation and movement in dangerous environments \cite{Search} has raised much interest
in terms of simulation studies over the last decade \cite{Virtual} due to the enormous losses in population with the increasing frequency and intensity of both man made or natural disasters. One
of the important application areas is emergency management where flow-based models \cite{wang2008modeling,pizzileo2010new}, queueing models \cite{lino2011tuning,desmet2013graph} and potential-maintenance models \cite{tseng2006wireless,chen2011distributed} are used to mimic an evacuation process and seek optimal solutions. Flow-based models imitate the evacuation planning problem as a minimum cost network flow problem. Evacuees are considered as continuous flows and capacity of nodes and edges is reflected by the restriction of flows. The graph model of the hazardous environment is extended to a time-expanded network by duplicating the original network for each discrete time unit and linear programming algorithms are used to solve optimal solutions. These approaches can achieve optimal solutions but do not consider the spreading of hazard. Queueing models are commonly used to analyse the time latency on routes and redirect evacuees accordingly. For instance, Ref. \cite{desmet2013graph} presents a computationally efficient algorithm based
on product form networks  \cite{gelenbe1987introduction} to predict the evacuation time with respect to average arrival and departure rates at each observation point. Potential-maintenance models treat an emergency environment as an ``artificial potential field'' and assign each on-site sensor with a potential value in accordance with distance to exits, distance to hazardous zone, etc. Evacuation routes are generated along sensors with higher potential value to sensors with lower potential value.

Such abstract models may be unrealistic so that simulators such as EVACNET+ \cite{kisko1985evacnet} have been proposed to provide an effective approach to evaluate routing algorithms. In recent years, the advancement in multi-agent technology has induced the emergence of realistic simulators \cite{kitano2001robocup,balasubramanian2006drillsim,dimakis2010distributed} which feature the self-observation and intra-competition of actors. This approach offers a more realistic approximation of hazard physical environments. Accompanying the maturity of simulation tools and the development of ICT technology as well as the reducing cost of sensors, enormous wireless sensor network based emergency response systems have been proposed. For instance, Ref. \cite{filippoupolitis2012emergency} proposes a distributed emergency navigation system that consists of sensor nodes (SNs) and decision nodes (DNs). The ``effective length'' of a route takes safety level and physical distance into account and Dijkstra's shortest path algorithm is used by DNs to calculate optimal paths for evacuees. Owing to the physical vulnerability of on-site electronic devices in hazardous environments, Ref. \cite{gelenbe2012wireless} replaces DNs with pocket wireless devices carried by evacuees. Rather than synchronising among all nodes in the system, hazard information are disseminated among evacuees within a certain communication range with the aid of opportunistic communications. Portable devices employ Dijkstra's algorithm to calculate shortest paths based on the received emergency messages.

However, wireless sensor network based emergency navigation systems suffer from inherent disadvantages such as limited computing capability, restrained battery power and restricted storage capacity. Hence, it is difficult for this type of architecture to provide optimal solutions in a timely fashion. Via using static or mobile sensors as thin client and offloading intensive computations to remote servers, cloud enabled emergency navigation systems have the potential to revolutionise this field. By leveraging existing public cloud services such as social network sites, Ref. \cite{li2011community,greer2012personal} present emergency warning systems to gather and disperse multi-media emergency information among users. Ref. \cite{ahn2011rescueme,gelenbe2014emergency} employ built-in camera on the smart phones to take snapshots and upload to servers to identify the location of evacuees. Based on the extracted data, the cloud-based emergency navigation system can provide appropriate paths for civilians.

By taking advantage of ``unlimited'' computing power, a few studies have devoted to realise faster-than-real-time prediction. Ref. \cite{han2010firegrid} presents a novel e-infrastructure to infer the spreading of hazard based on predictive models and living sensory data in a faster-than-real-time manner. However, to the best of our knowledge, there is no research using faster-than-real-time simulation to optimise an evacuation process.

\section{The Cloud-Enabled Emergency Navigation System}
\label{system}

\begin{figure}[h!]
\centering
\includegraphics[width=0.4\textwidth]{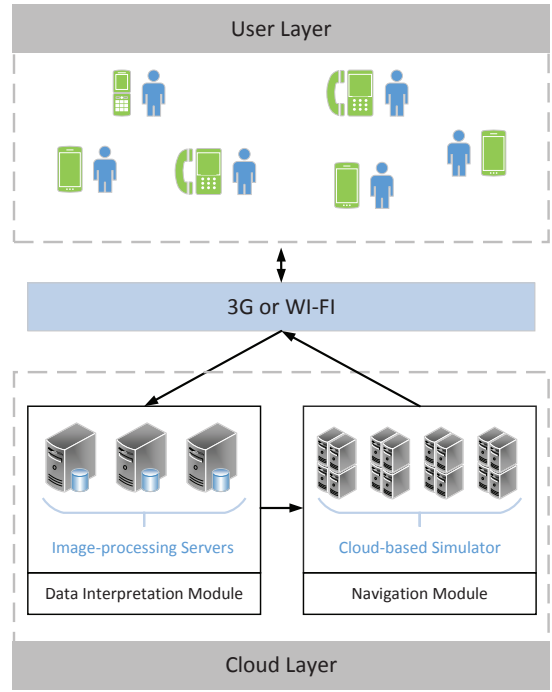}
\caption{The architecture of the cloud-enabled system.}
\label{fig: systemstructure}
\end{figure}

The proposed cloud-enabled emergency navigation system consists of user layer and cloud layer as shown in Fig. \ref{fig: systemstructure}. The user layer is deployed on handhold devices and is responsible for collecting on-site information and exchanging data with cloud servers while the cloud layer harnesses massive inter-networked computers on server-side to provide a platform for the system. When an disaster occurs, evacuees will use smart phones to take snapshots and upload to the cloud layer to identify their initial position. The cloud layer contains a data interpretation module and a navigation simulation module. The data interpretation module possesses a number of image processing servers that can extract landmarks from the uploaded photos and match them with pre-stored images to determine the initial location of evacuees. The navigation module maintains a simulated environment for the conduction of emergency navigation algorithm. The simulator is a distributed simulator that can dispatch workload to a large amount of inter-connected high performance servers. Each server is associated with one significant position such as doorway in the built environment and maintains information observed by traversed evacuees. Rather than synchronising collected data among all servers periodically, servers will send packets to gather information from other servers. The user layer and the cloud layer can communicate via diverse gateways such as Universal Mobile Telecommunications System (UMTS) or Internet.

\section{Routing Algorithms}
\label{algorithms}

Linear programming approaches, which consider evacuation route planning as a network flow problem \cite{jarvis1982note}, can solve optimal routes for evacuees. The common procedures of a linear programming approach are as follows. First, it predicts the upper bound of evacuation time and converts the network graph to a time-expanded network by duplicating the original network for each discrete time unit. Second, it employs a linear programming method to solve the time-expanded network. However, to achieve shortest time to exit, evacuees must accurately follow the suggested paths and reach every node on schedule and may even wait certain time at a node to avoid congestion. This is impractical in a real evacuation process. Moreover, these approaches suffer from high computational complexity because the time-expanded network will contain at least $(N+1)T$ nodes for a graph with $N$ nodes and an upper bound of evacuation time $T$.

Due to the drawbacks of linear programming approaches, we propose a relatively computationally efficient and more robust heuristic algorithm to guide evacuees. This emergency navigation algorithm is deployed in the navigation simulation module. This simulator can perform faster-than-real-time simulation with respect to the initial situation uploaded by smart phones. According to the simulation results, the routes of the fatalities will be re-assigned to improve the survival rate. In this system, two routing algorithms are employed: a variant of Dijkstra's algorithm and Cognitive Packet Network based algorithm. The original Dijkstra's algorithm, which does not take congestion level of a path into account, can obtain optimal solutions in relatively low occupancy rates. However, it can induce serious fatalities in densely-populated environments because it tends to guide all evacuees to the shortest path can cause severe congestion. On the other hand, Cognitive Packet Network based algorithm which can reach the performance of Dijkstra's algorithm in low population densities and achieve better survival rates in high population densities. Hence, the Cognitive Packet Network based algorithm with time metric \cite{bigelenbe2014routing}, which achieves the overall best performance, is selected as the basis algorithm and a variant of Dijkstra's algorithm which considers congestion level of a path is used to reallocate paths for fatalities in the previous simulation. The detailed procedures are shown in Pseudocode \ref{pseudo: routingalgorithm}.

\begin{algorithm}[ht!]
\floatname{algorithm}{Pseudocode}
\caption{The procedures of directing evacuees. EMS, emergency management system. CPNST, Cognitive Packet Network based algorithm with time metric.}
\label{pseudo: routingalgorithm}
\begin{algorithmic}[1]
\State When a hazard is detected, EMS alerts every evacuee in the building.
\State Evacuees upload snapshots to inform their initial position.
\State Based on the uploaded information, a faster-than-real-time simulation is conducted and CPNST is employed to generate routes for evacuees.
\State For all the fatalities in the simulation, a time-dependent Dijkstra's algorithm is used to reassign routes for perished civilians.
\State Send calculated routes to evacuees.
\State Civilians use the received routes to egress in a source-routed manner.
\end{algorithmic}
\end{algorithm}

In the following two subsections, details of the involved two algorithms are introduced.

\subsection{Cognitive Packet Network}

The Cognitive Packet Network (CPN) \cite{gelenbe2009steps} is a quality of service (QoS) driven protocol that was originally proposed for route-finding in large-scale packet networks as well as tailoring diverse quality of service (QoS) for end users in multimedia networks. CPN was arguably the first software-defined network (SDN) in the world and can be readily deployed on various hardware devices. A CPN is composed of CPN nodes in which a Random Neural Network (RNN) \cite{gelenbe1990stability,Synchronized,Natural} is installed as a decision-making algorithm, and a mailbox (MB) is maintained to store observed information. CPN contains three types of packets, smart packets (SPs), acknowledgements (ACKs) and dumb packets (DPs). SPs are responsible for path-finding and information-gathering. Each SP can carry a cognitive map as well as executable code and control its own behaviours. The search direction of SPs is determined by either RNN maintained at each CPN node or a random walk behaviour controlled by ``drift parameter''. Drift parameter is defined as the possibility for a SP to choose the next hop at random over the RNN's advice. When a SP reaches the destination, it will generate an ACK to bring back all the collected sensory information. DPs carry the payloads and always follow the top-ranked route in the mailbox discovered by SPs. In the context of emergency navigation, evacuees are considered as DPs.

CPN is naturally a distributed system and can be readily integrated with cloud servers. Each CPN node is installed on a cloud server and SPs are used to gather interested information from other servers. Furthermore, since each server also represents a significant position in the built environment, the routes discovered by SPs among servers can be directly used as the paths for evacuees.

In the previous work, we borrowed the concept of CPN and developed several sensible routing metrics \cite{Sensible} to guide diverse evacuees with specific QoS requirement \cite{bi2014routing}. Among these algorithms, CPN with time metric (CPNST), which pursues routes with shortest time to exits via avoiding congestion, achieves the overall best performance. In this paper, we use CPNST as the basis algorithm to guide evacuees. However, certain evacuees may perish when using CPNST since this algorithm takes the risk to traverse potential hazard areas in order to reduce the evacuation time. Hence, we also develop a time-dependent Dijkstra's algorithm to customise paths for all the perished evacuees caused by using CPNST.

\subsection{Time-dependent Dijkstra's Algorithm}

Dijkstra's shortest path algorithm and its variants are often used in path finding and have been introduced to wireless sensor network based emergency navigation systems. However, since Dijkstra's algorithm needs to visit every sensor node in the network to solve the optimal solution, some problems may occur in the WSN based emergency response systems such as direction oscillation problem \cite{chen2008load} caused by communication latency, signal interference between nodes and high energy utilisation during information exchange. These issues can be avoided in our algorithm because the time-dependent Dijkstra's algorithm is conducted on the server side. The proposed adaptation of Dijkstra's algorithm follows the same fashion as the classical Dijkstra's shortest path algorithm except that we replace the distance-dependent metric with a time-dependent metric. The novel metric is shown in (\ref{equ: timemetric}).
\begin{equation}
\label{equ: timemetric}
T(\pi(i), \pi(i+1)) = \frac{E^e(\pi(i), \pi(i+1))}{V_s} + t^q_c(\pi(i))
\end{equation}
where $\pi$ represents a particular path and $\pi(i)$ depicts the $i$-th node on the path $\pi$. $E^e(\pi(i), \pi(i+1))$ is the effective length \cite{filippoupolitis2009distributed} of the edge between node $\pi(i)$ and $\pi(i+1)$. It is a combined metric which takes path physical length and safety into consideration. If an edge is not affected by hazard, its effective length equals to the physical length. Otherwise, the effective length is much larger than physical length to ensure a path exposed to fire will not be selected if any safe path exists. Term $V_s$ represents the average speed of civilians. Term $t^q_c(\pi(i))$ depicts the queueing time after an evacuee reaches node $\pi(i)$.

The queueing time $t^q_c(\pi(i))$ at node $\pi(i)$ can be calculated by the number of evacuees at the node when the evacuee arrives. It can be readily obtained from the simulation by recording the number of evacuees at all instants. We assume that each evacuee will cost one second to determine the next direction when arriving at a node and evacuees will leave a node one by one.
\begin{equation}
\label{equ: timecostatnode}
t^q_c(\pi(i)) = \frac{N_q^c}{d_{\pi(i)}}
\end{equation}
where $N_q^c$ is the number of queued civilians when the evacuee arrives and $d_{\pi(i)}$ is the departure rate of node $\pi(i)$.

The detailed procedures of the time-dependent algorithm are shown in Pseudocode \ref{pseudo: dijkstra}:

\begin{algorithm}[ht!]
\floatname{algorithm}{Pseudocode}
\caption{The procedures of the time-dependent Dijkstra's algorithm. CPNST, Cognitive Packet Network based algorithm with time metric.}
\label{pseudo: dijkstra}
\begin{algorithmic}[1]
\State Conduct a simulation using CPNST algorithm.
\ForAll {the nodes in the building model}
\State Record the number of evacuees $N_q^c(T)$, the hazard intensity $H^c(T)$ at all instants $T = 0, 1, 2 \ldots$
\EndFor
\State Sort the perished evacuees in the simulation based on the distance between the initial position and the nearest exit in ascending order.
\ForAll {the perished evacuees in the simulation}
\State Perform time-dependent Dijkstra's algorithm to calculate the path with the shortest time to exits.
\EndFor
\end{algorithmic}
\end{algorithm}

\section{Simulation Model and Assumptions}
\label{simulation}

\begin{figure}[!ht]
\centering
\includegraphics[width=0.4\textwidth]{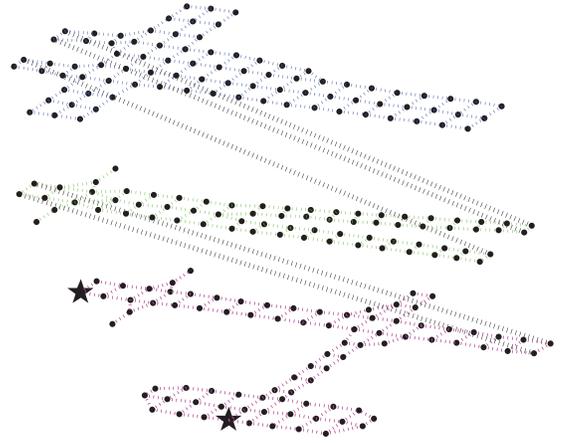}
\caption{Graph representation of the building model. The two black stars on the ground floor mark the position of the building's exits.}
\label{fig: graph3D}
\end{figure}

We use the Distributed Building Evacuation Simulator (DBES) to evaluate the effectiveness of the proposed algorithm in fire-related scenarios. DBES is a java-based multi-agent discrete event simulator where evacuees are modeled as agents and can interact with other agents and make decisions based on their own observations. In the simulator, physical areas are depicted by a graph model: vertices represent points of interest (PoI) such as locations where evacuees may congregate; edges depict the physical routes between PoI. Our experiments are conducted in a canary wharf shopping mall as shown in Fig. \ref{fig: graph3D}. We assume that no sensors are pre-installed in the building and evacuees can identify their position by taking snapshots at PoI where visual landmarks are presented and then uploading photos to image-processing servers. Evacuees are hypothesised to carry smart phones with camera and are initially distributed in the building at random.

The fire spreading model is based on \cite{elms1984modeling} and details can be seen in \cite{filippoupolitis2008emergency}. Each vertex possesses a hazard intensity which is affected by its previous hazard intensity as well as neighbouring vertices. The hazard intensity of an edge is presented by the average value of two endpoints. Every edge also has a specific hazard spreading rate in terms of the physical link it represents such as corridor, doorway or staircase. Fire randomly propagates along edges and therefore the spreading of hazard will be different for each simulation run.

The core of the proposed algorithm is the simulation process which can discover evacuees with a high potential to perish and re-route them. Hence, to realise faster-than-real-time simulation, we assume a data center with numerous  inter-connected servers are used. Each server is associated with a vertex in the building model. Hence, in our case, 243 servers are involved in conducting the faster-than-real-time simulation. The CPU frequency of each server is 3.4 $GHz$ and the Instructions per cycle (IPC) of the CPU is set to 1.

The elapsed time of a whole simulation is evaluated with regard to the CPU frequency of servers and the recorded CPU cycles during the simulation. The CPU cycles are calculated by an analogy of the code-transformation algorithm introduced in Ref. \cite{shnayder2004simulating}. Detailed procedures of this algorithm are as follows:

\begin{enumerate}
  \item Convert the java codes of the algorithms to simple CPU instructions;
  \item Convert involved decimal numbers in the algorithms to binary numbers;
  \item Perform bitwise operations and sum the number of instructions used in computation;
  \item Transform the number of instructions to CPU cycles counts in accordance with the Instructions per Cycle;
  \item Convert CPU cycles counts to CPU active time with regard to CPU frequency;
\end{enumerate}

\section{Results and Discussion}
\label{results}

To evaluate the performance of the proposed algorithm, we simulate a fire-related disaster in the aforementioned building model. The proposed algorithm which combines CPN with time metric and time-dependent Dijkstra' algorithm (CPNST\&TD) is conducted under variant occupancy rates (30, 60, 90 and 120 evacuees). Dijkstra's shortest path algorithm (DSP) and original CPN with time metric (CPNST) are performed for comparison purposes.

\begin{figure}[!ht]
\centering
\includegraphics[width=0.5\textwidth]{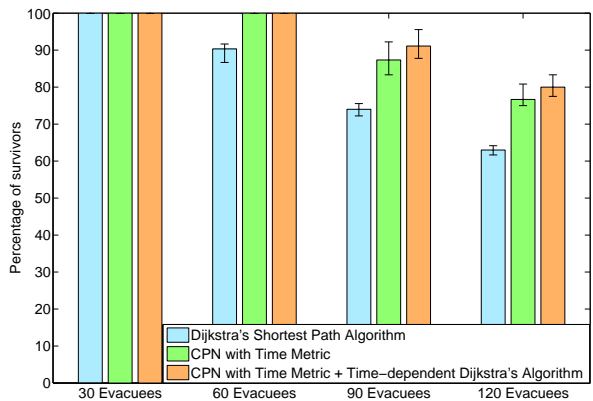}
\caption{Percentage of survivors for each scenario. The results are the average of five randomized simulation runs, and error bars show the min/max result in any of the five simulation runs.}
\label{fig: numberofSurvivors}
\end{figure}

As can be seen in Fig. \ref{fig: numberofSurvivors}, although DSP can always provide shortest path for each individual, CPNST achieves better performance than it. This is because CPNST can ease congestion by dynamically redirecting evacuees to uncrowded paths while DSP tends to guide all evacuees to the shortest main channel and induces severe jamming. CPNST\&TD, which is based on CPNST, achieves the best performance in all four scenarios despite of varying occupancy rates. The reason is that CPNST is sensitive to the changes of route conditions such as congestion level or hazard intensity. Although this is an advantage for evacuees relatively nearer to exits (first and second floors) as evacuees can effectively avoid congested path, certain distant civilians on the third floor may suffer from direction oscillation at the beginning if several candidate paths are equally good. Moreover, unlike CPNST\&TD which can obtain the hazard intensity of each vertex at any time instant by performing a faster-than-real-time simulation beforehand, CPNST may take the risk to traverse potential hazard areas in order to reduce the evacuation time. As a result, certain evacuees may significantly prolong the evacuation time when they are suddenly blocked by fire and have to select a detour path. To the contrary, CPNST\&TD can the avoid above problem as it reassigns paths for perished evacuees in the simulation.

\begin{figure}[!ht]
\centering
\includegraphics[width=0.5\textwidth]{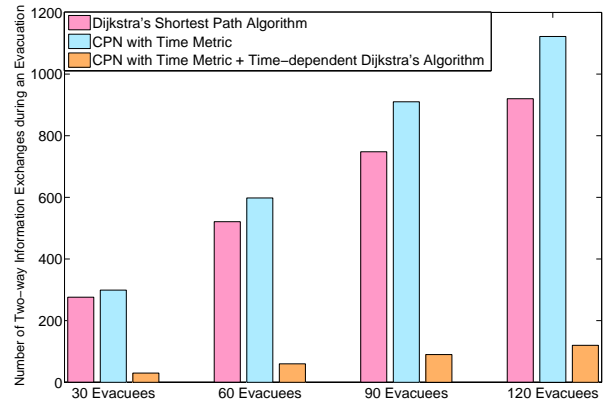}
\caption{Number of two-way information exchanges between smart phones and the Cloud during an evacuation process for diverse occupancy rates.}
\label{fig: numberofinformationexchanges}
\end{figure}

Because it is not cost effective to maintain an over-provisioned communication infrastructure for these infrequent emergency events, tsunami of data can be caused by the overwhelmed information exchanges between cloud servers and smart phones. Fig. \ref{fig: numberofinformationexchanges} shows the number of two-way information exchanges between smart phones and cloud servers during an evacuation process. One two-way information exchange is defined as the process of an evacuee uploading a snapshot and gaining its location and suggested path from the Cloud. Owing to the effect of highly dynamic hazardous environment such as spreading of hazard and abrupt congestion, the suggested paths of DSP and CPNST which are solved based on living sensory data need to be updated periodically. Hence, in those two algorithms, evacuees need to upload sensory data such as congestion or hazard information and download advices from the Cloud frequently. On the other hand, CPNST\&TD, which produces solutions based on simulation results, only concerns the initial conditions. Hence, CPNST\&TD only needs to exchange information with the Cloud at the beginning of an evacuation process. As expected, CPNST\&TD achieves the lowest number of data exchanges with cloud servers. This is because CPNST\&TD can send final paths to evacuees when a disaster breaks out rather than making periodically decisions based on living sensory data. CPNST exchanges information more frequently than DSP because evacuees using CPNST do not follow the shortest path. Hence, evacuees will traverse more landmarks and upload more photos to the Cloud.

The elapsed time of the proposed algorithm, which experiences a simulation process and a path re-assigning process, is evaluated in the aforementioned cloud environment with 243 servers. The elapsed time for a 136 seconds long evacuation process is 2.63 seconds and the recorded total CPU cycles are $1.92*10^{12}$. The time cost for the path re-assigning process is 0.39 seconds and the CPU cycle consumption is $3.74*10^{10}$. Hence, the total elapsed time for performing the proposed algorithm is 3.02 seconds, which confirms that our algorithm can provide appropriate paths for evacuees in a faster-than-real-time manner.

\section{Conclusions}
\label{conclusion}

In this paper we propose a simulation-based routing algorithm to increase the survival rate of an evacuation process. Contrary to the traditional algorithms which normally create a delayed feedback loop between living sensory data and routing decisions, the proposed routing algorithm uses a cloud based simulator to predict the result of CPNST and re-calculate optimal paths for perished civilians in the simulation. The experimental results indicates that the presented algorithm achieves improved survival rates. Moreover, since a fire model is used to predict the spreading of hazard, the simulator can calculate desired routes only based on the initial distribution of evacuees. Therefore, the data tsunami, which is created by the massive information uploading and downloading shortly after the onset of an emergency event, is avoided.



%
\bibliographystyle{IEEEtran}
\bibliography{simulationbasedalgorithm}

\end{document}